# Fabry-Perot Lasers as Enablers for Parallel Reservoir Computing

Adonis Bogris, Charis Mesaritakis, Stavros Deligiannidis, Pu Li

*Abstract*— We introduce the use of Fabry-Perot (FP) lasers as potential neuromorphic computing machines with parallel processing capabilities. With the use of optical injection between a master FP laser and a slave FP laser under feedback we demonstrate the potential for scaling up the processing power at longitudinal mode granularity and perform real-time processing for signal equalization in 25 Gbaud intensity modulation direct detection optical communication systems. We demonstrate the improvement of classification performance as the number of modes multiplies the number of virtual nodes and offers the capability of simultaneous processing of arbitrary data streams. Extensive numerical simulations show that up to 8 longitudinal modes in typical Fabry-Perot lasers can be leveraged to enhance classification performance.

## I. INTRODUCTION

Artificial neural networks (ANNs), and machine learning in general, are becoming ubiquitous for an impressively large number of applications. This fact has brought ANNs into the focus of research in computer science and several topics such as machine/deep learning have conquered almost every domain of scientific areas related to big data, signal processing and everyday life as well. Traditional von Neumann computer architectures, although efficient for mathematical calculations, face difficulties when highly complex or abstract computational tasks such as speech recognition or facial recognition are targeted. On the other hand, human brains operate in a different way and seem to be most appropriate for this kind of tasks. This widely accepted fact has led the scientific community to investigate and implement hardware based neuromorphic computing architectures, in other words, hardware platforms that can mimic human brain functions. There have been numerous demonstrations of different hardware neuromorphic computing paradigms which emerge as interesting computing systems provided that they have the potential to overcome conventional computers in terms of speed and/or energy efficiency. Different hardware neuromorphic computing paradigms based on solid-state platforms, mechanical devices, memristor arrays, optical systems and spintronic devices have been reported [1-5]. The photonic neuromorphic processors can be very promising computing machines as they rely on ultrafast physical mechanisms and have been mostly used in reservoir computing (RC). The concept of RC offers a framework to exploit the transient dynamics of a recurrent neural network (RNN) that are ideal for time-evolving problems and at the same time it enables easy training as the whole training process

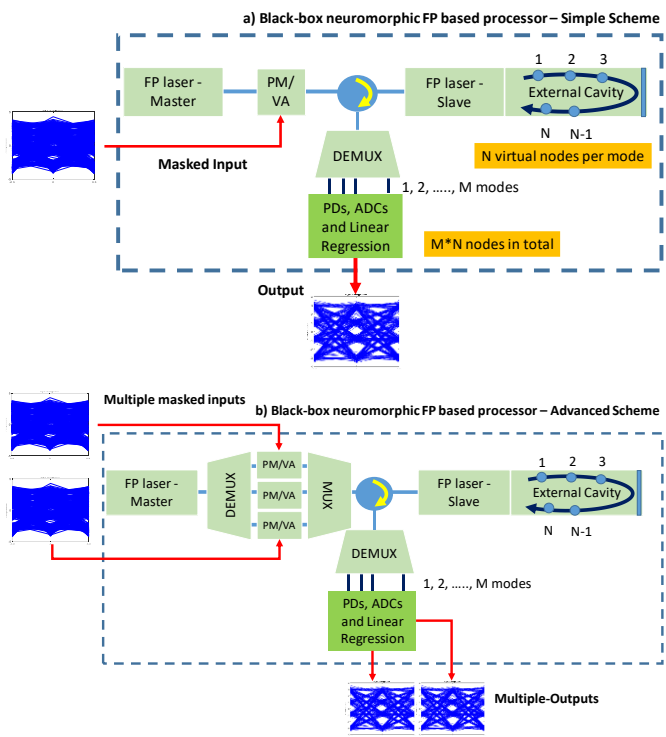

Fig. 1. Neuromorphic processor based on FP lasers, a) The simple architecture, b) the advanced architecture offering multi-signal processing in optical communications.

Manuscript received XXXX. This work has been partially funded by the H2020 project NEoteRIC (871330). This project has also received funding from the Hellenic Foundation for Research and Innovation (HFRI) and the General Secretariat for Research and Technology (GSRT), under grant agreement No 2247. Adonis Bogris and Stavros Deligiannidis are with the Department of Informatics and Computer Engineering, University of West Attica, Aghiou Spiridonos, Egaleo, 12243, Athens, Greece (e-mail: sdeligiannid@uniwa.gr, abogris@uniwa.gr).

Charis Mesaritakis is with the Department of Information & Communication Systems Engineering, University of the Aegean, 2 Palama & Gorgyras St., 83200, Karlovassi Samos, Greece (e-mail: cmesar@aegean.gr).
Pu Li is with Key Laboratory of Advanced Transducers and Intelligent Control System, Ministry of Education, Taiyuan University of Technology, Taiyuan 030024, China (e-mail: lipu8603@126.com).

TABLE I
NUMERICAL MODEL PARAMETERS

| Symbol | Parameter | Value |
|---|---|---|
| $g$ | Differential gain parameter | $1.2 \times 10^{-8}$ ps$^{-1}$ |
| $s$ | Gain saturation coefficient | $5 \times 10^{-7}$ |
| $\beta$ | Spontaneous emission rate | $1.5 \times 10^{-10}$ ps$^{-1}$ |
| $t_n$ | Carrier lifetime | 2 ns |
| $N_0$ | Transparency Carrier Number | $1.5 \times 10^{8}$ |
| $a$ | Linewidth enhancement factor parameter | 3 |
| $T$ | External cavity round-trip time | From 240 ps to 1 ns |
| $t_{ph}$ | Photon line-time | 2 ps |
| $\omega_0$ | Central oscillation frequency | $1.216 \times 10^{15}$ rad/sec |
| $\Delta f_L$ | slave laser FSR - $FSR_s$ | 125 GHz |
| $\Delta f_g$ | Gain bandwifth | 10 THz |
| $\theta$ | Node separation | 20 ps |
| $M$ | Number of longitudinal modes | 8 |
| $k_{inj}$ | Injection parameter | 0.75 |
| $k_f$ | Optical feedback parameter | 0.01 |

relies on a single output layer of nodes with trainable weights using a linear regression algorithm [6]. Up to date, there have been many demonstrations of optical RC exploiting two main schemes: the spatially distributed systems, connecting physical nodes made of nonlinear oscillators, as well as delay-based approaches. The concept of delay line-based RC using only a single nonlinear node with delayed feedback was introduced in [7, 8] as a promising way to minimize the hardware complexity in photonic systems. A number of classification, prediction, and system modelling tasks have been demonstrated with the use of delay-based optical RC, with excellent performance in speech recognition, chaotic time series prediction, nonlinear channel equalization amongst other [8-10]. All-optical RC based on delay systems, that is lasers subjected to optical feedback, have the potential to be integrated in a photonic chip. This potential has been demonstrated by Harkoe et al. [11] and it paves the way for faster (since processing speed is directly related to the RC delay time), power efficient and more compact RC systems.

Recent research in optical RC systems has shown that the processing power of the system can be enhanced by higher modulation bandwidth schemes assisted by optical injection techniques [12] and/or by using multiple lasers [13], the dual polarization dynamics of lasers of VCSELs [14] and modal degree of freedom in dual-mode lasers [15]. Similarly, the simultaneous processing of two independent tasks has been shown in a semiconductor ring laser emitting in two directional optical modes [16]. In this paper we present the potential of off-the-shelf Fabry-Perot lasers in serving as ultrafast processors that take advantage of the numerous longitudinal modes in order to perform many tasks in parallel or to enhance the performance in a single task by increasing the number of virtual nodes offered by multiple longitudinal modes. Exploiting both injection locking and optical feedback we perform reservoir computing processing that employees up to 8 longitudinal modes in typical FP diode lasers. Through numerical simulations, the paper shows the practicality and versatility of the FP-based RC scheme in compensating for the accumulated dispersion affecting 25 Gbaud PAM-4 signals after 50 km of transmission in single mode fiber. The paper shows the potential of Fabry-Perot based RCs for real-time operation approaching 10 Gsa/s computation speed and monolithic integration. Section II describes the main features of the system and the modeling approach.

## II. RESERVOIR COMPUTING SYSTEM BASED ON FABRY-PEROT LASERS

### A. Basic concept

The black-box neuromorphic processor relying on Fabry-Perot lasers is depicted in fig. 1 in two variants. Both variants involve a master FP laser that is phase modulated (PM) with the properly masked signal to be processed – in our case, the PAM-4 signal that has experienced transmission along 50 km of fiber. The modulated multi-mode signal is injected into a second FP laser with injected power adjusted by a variable attenuator (VA). The second laser, the so-called slave, is subjected to feedback with the use of a short external cavity. Short cavities are considered as the paper focuses on solutions that can be monolithic integrated. The output of the slave laser is sent to a wavelength demultiplexer (DEMUX) that drives each mode to a separate photodiode and then to an analog to digital converter (ADC) unit. The $M$ outputs, where $M$ stands for the number of modes taking place in this process, either increase by a factor of $M$ the virtual nodes before linear regression, or they can be grouped in as many sets as the number of signals that have to be processed concurrently. The latter can be solely served by the advanced scheme (fig. 1b) which shows that each FP master mode can be modulated and controlled individually so as to accommodate and process more than one signals at the same time.

In delay-based reservoir computing systems, the nodes are virtually distributed in the temporal domain in order to define multiple virtual nodes within the feedback loop characterized by a round-trip delay equal to $T$. Towards this goal, the input data are first pre-processed as follows: first, the input digitized symbols are sampled and held for a period of $T$ such that the injected symbol is spread across the duration of a round-trip in the slave external cavity. If the symbol rate of the input signal is $T_s$, then the ratio $T/T_s$ shows the time-stretching factor to accommodate each symbol in the external cavity time. The shorter the stretching factor is, the smaller the speed penalty that characterizes the system. In [17] simulations on 25 Gbaud PAM-4 processing showed that speed penalty is 40 for a BER performance below $10^{-3}$ when both master and slave are single mode lasers. That means that the RC system can not really operate in real-time in high speed telecom applications as it has to store 40 "packets" before having finished the processing of a single packet. Even if a buffer of infinite capacity is considered, the large latency that is attributed to signal storing and slow processing will lead to time-outs, packets retransmissions and throughput reduction in modern TCP/IP networks. Only slowing the input stream at the level of RC dynamics is a viable solution which however severely affects throughput.

The number of virtual nodes in the interval $T$ is $N_v=T/\theta$ where $\theta$ is the well-known node separation and must have a value relevant to the dynamical behavior of the slave laser – usually multi-GHz dynamics can be achieved - and must comply with specifications of state of the art ADCs (thus not lower than 20 ps). The masking process is the second pre-processing stage needed to take place at the input. By multiplying each symbol that has been stretched to $T$ duration with a set of $N_v$ random values that are periodically repeated every after $T$, the input weights are applied. The reason we use phase modulation instead of intensity modulation is to enhance the dynamical behavior of the slave laser in order to reduce $\theta$ at values that are only limited by state of the art ADC technology. If $N_v$ is small, the processing power reduces although the speed penalty is relaxed. Thus, leveraging more than one longitudinal modes one can boost the processing power and reduce the speed penalty at the same time [15]. If $M$ longitudinal modes are employed, then the total number of virtual nodes will become $M \times N_v$.

### B. The FP master-slave laser model

In our analysis typical FP semiconductor lasers are considered. The lasers are modeled based on the well-known rate equations for the complex slowly varying amplitude of the electrical field $E_{s,k}$ of $k$-th mode and the carrier number inside the cavity $N$ given below [18]. For the sake of simplicity and without loss of generality, only the equations for the slave laser are provided below and we consider identical intrinsic parameters for both lasers. We have tested completely different parameters for master laser than those provided in Table I and the results did not change for the same optical injection strength and frequency detuning between master and slave lasers.

$$\dot{E}_{s,k}(t) = \frac{1+i\alpha}{2}\left[G_{s,k}(t) - \frac{1}{t_{ph}}\right]E_{s,k}(t) + \frac{k_f}{t_{rt}}E_{s,k}(t-T)e^{i\omega_k T} +$$

$$+ \frac{k_{inj}}{t_{rt}}E_{m,k}(t)e^{-i\Delta\omega_k t} + \sqrt{2\beta N(t)}\xi(t)$$

$$\dot{N}(t) = \frac{I}{q} - \frac{1}{t_n}N(t) - \sum_{k=1-M/2}^{M/2}G_{s,k}|E_{s,k}(t)|^2$$

$$G_{s,k}(t) = \frac{g[N(t)-N_0]}{1+s\sum_{k=1-M/2}^{M/2}|E_{s,k}(t)|^2}\left[1-\left(k\frac{\Delta f_L}{\Delta f_g}\right)^2\right]$$

(1)

In (1), $a$ is the linewidth enhancement factor, $g$ is the differential gain parameter, $s$ is the gain saturation coefficient, $t_{ph}$ is the photon lifetime, $t_n$ is the carrier lifetime, $N_0$ is the carrier number at transparency $G_{s,k}$ is the modal gain per unit time at $k_{th}$ mode and $I$ is the bias current. Moreover, $\beta$ is the spontaneous emission factor and $\xi(t)$ the spontaneous emission process (Langevin forces) modelled as a complex Gaussian process of zero mean and correlation $\langle\xi(t)\xi^*(u)\rangle = 2\delta(t-u)$. $E_{m,k}$ corresponds to the $k$-th mode of master laser modulated in phase with the term $m(t)\cdot d(t)$ where $m(t)$ is the mask signal and $d(t)$ is the distorted PAM-4 data. The mask signal $m(t)$ is the random connectivity matrix at the input with $N_v$ values that take values from 0 to 1 similarly to [17]. The phase modulation $m(t)\cdot d(t)$ term takes values from 0 to $2\pi$. The term $\Delta\omega_k$ is the angular frequency detuning between the $k$-th mode of master and slave lasers. $\Delta f_L$ corresponds to the free spectral range (FSR) of the slave laser which is equal to $\Delta f_L = 1/t_{rt}$ where $t_{rt}$ corresponds to the laser cavity round-trip time and $\Delta f_g$ is the gain bandwidth of the diode laser, usually around 10 THz. Regarding injection and feedback terms, $k_f$ is the parameter that corresponds to the amount of optical feedback in slave external cavity and $k_{inj}$ corresponds to the percentage of master laser amplitude injected to slave corresponding mode. Although the aforementioned model does not take into account spatial hole burning effects and four-wave mixing between adjacent modes, it has been proved a quite reliable model when commercially available Fabry-Perot diode lasers with FSR in the range of 120 GHz are considered, as in our case, for a wide range of applications [19, 20]. Obviously if longer cavities were considered that translate to FSR values below 50 GHz, then coherent mixing between the modal fields could not be neglected and more sophisticated but also more complex and time-consuming modeling would be necessary [21]. The values of the most critical parameters are summarized in Table I. For these values, typical lasers with relative intensity noise in the order of -140 to -160 dB/Hz are considered. The external cavity round-trip is decided to be below 1 ns, to assume that the whole system can be monolithic integrated [22]. Miniaturization perspective is discussed in the next section.

### C. Data training and performance evaluation in PAM-4 transmission

In this paper we benchmark the FP based RC as a dispersion compensator in intensity modulation direct detection optical communication systems. Following the approach of [17] we numerically simulate the transmission of PAM-4 25 Gbaud signals over 50 km of typical single mode fibers and then we use the RC system in order to mitigate dispersion effects and improve the bit-error rate (BER) performance. Signal propagation in our model is derived with split-step Fourier method governed by Manakov equations [23]. We also take into account thermal and shot noise at the photodetector. The distorted signal is inserted in the reservoir using random input weights with the masking process described above. At the output, we sample the signal every after $\theta$ to get the response of each virtual node, then we quantize with 8-bit resolution and then we linearly combine their responses to identify the initially transmitted PAM-4 signal. The optimal weights of the linear summation among the outputs of the nodes are derived with the use of ridge regression algorithm and the efficiency of RC processing is evaluated in the validation process by estimating BER through error counting in symbol by symbol comparison between the RC output and the expected value per symbol. Training is based on 60000 symbols and testing and validation in 20000 symbols. The classifier is not trained per symbol but using a batch of consecutive symbols that better resemble the memory of the fiber channel [17]. This batch contains 21 symbols in our simulations. For such a number of trained symbols the number of weights of the linear classifier at the output that are needed are $M$ x $N_v$ x 21. The numerical results are presented in section III.

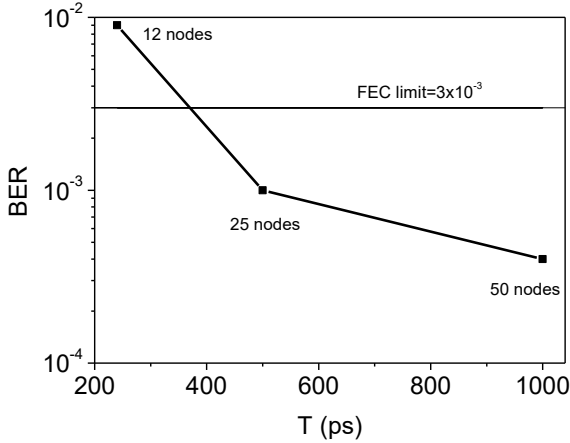

Fig. 2. BER performance as a function of $T$ in single mode lasers. For $T > 500$ ps ($N_v>25$), BER becomes lower than $10^{-3}$ which is the FEC limit. Without compensation, BER=0.06

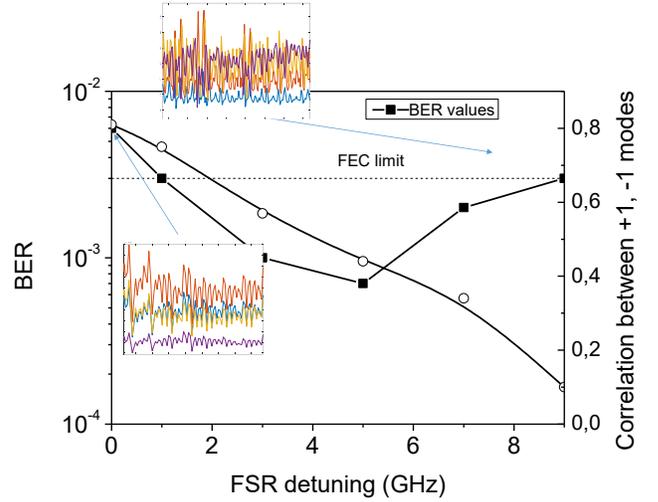

Fig. 3. BER performance as a function of the FSR detuning between the two lasers. In the right vertical axis, the correlation coefficient between +1 and -1 modes is also depicted. It becomes clear that worse BER performance occurs when FSR difference is almost zero. In this regime, higher correlation between +1, -1 modes is found.

## III. RESULTS AND DISCUSSION

### A. Evaluation of the simple RC scheme

First we study the simple scheme depicted in fig. 1a. This scheme is simpler as it employs only one modulator, thus all longitudinal modes of master FP will launch the same phase modulation (same data, same mask) into the slave laser modes. As already reported, the results concentrate on the BER performance of the distorted PAM-4 signal for different number of modes participating in the computing process. The BER of a 25 Gbaud PAM-4 signal that has propagated along 50 km of fiber is BER=0.06 even if the linear classifier is used right after photodetection and digitization. The external cavity delay of the slave laser is kept below $T=1$ ns and in this work we consider as minimum and targeted external cavity delay $T=240$ ps as our goal is to achieve miniaturization of the whole system ($T=240$ ps corresponds to approximately 1 cm long external cavity in III-V materials). The number of virtual nodes per mode are $N_v=T/\theta$ with $\theta=20$ ps and scales from 12 to 50. If one wants to increase the number of nodes, one has either to increase $T$ at expense of speed, or decrease $\theta$ which is not possible with current state of the art ADC technology.

Initially, we test the efficacy of the single-mode system, as a function of $T$. We have carried out an optimization analysis and found that the RC based on single-mode laser system operates very well when both lasers are biased with 30-35 mA (twice the threshold), $k_{inj}=0.75$ and $k_f=0.01$. The feedback phase has been kept constant to $\pi/4$ in all simulations and its variation does not seem to affect the performance. Their frequency detuning is set to 0. For these laser and injection/feedback parameters we derived dispersion compensation with the use of RC and we estimated BER through error counting. The results are depicted in fig. 2. It is apparent that the system has the ability to provide acceptable BER values, that is below $3\times10^{-3}$ forward error correction (FEC) limit, when $T$ exceeds approximately 400 ps. At this value, the speed penalty induced by the process is 10 ($T=400$ ps, $T_s=40$ ps). With the use of multi-mode lasers we envisage to further reduce this number.

The results to be presented in this paragraph consider multi-mode operation of the two lasers. Eight modes have been considered and $T=240$ ps. Thus, 96 virtual nodes can be provided and speed penalty reduces to 6. When multi-mode injection is attempted from master to slave and taking into account that all modes carry exactly the same phase modulation pattern (simple scheme, fig. 1a) and pass through the same VA, one has to identify what the optimal injection parameters should be and which degrees of freedom can be exploited to de-correlate the time sequences at the output modes of the slave laser. The de-correlation is necessary to enrich the dynamics offered by incorporating more than one modes. In other words, if all modes of slave laser exhibit exactly the same nonlinear response, then the benefit of combining the outputs of individual modes will be negligible. The only parameter that can differentiate the injection conditions from master to slave for each mode is master-slave frequency detuning. This can be realized if master and slave lasers exhibit different FSR values. Different FSR values mean that each master mode injects the corresponding slave mode with a different frequency detuning in terms of value and sign, thus nonlinear dynamics per mode can be different, although there is interdependence of inter-mode dynamics as all modes share the same carrier pool. Assuming that master injects slave with zero detuning at the central mode ($k=0$), then its detuning for $k$-th mode with $k$ taking values from -4 to 3 (eight modes in total) will be $k*(FSR_m-FSR_s)$. For instance, if $FSR_m=128$ GHz and $FSR_s=125$ GHz, then mode 1 of master enters mode 1 of slave with $\Delta\omega_1$ =+3 GHz x $2\pi$ detuning whilst mode -1 has a detuning of $\Delta\omega_{-1}$ =-3 GHz x $2\pi$. We studied BER performance as a function of $FSR_m$-$FSR_s$ and we provide the results in fig. 3. At the same figure we also depict the correlation coefficient evaluated when comparing the output power of mode 1 and mode -1. On one hand, the higher their correlation, the less the differentiation of the output power of slave modes. On the other hand, high FSR differences make less efficient or even impossible the injection locking between master and slave at higher order modes even if injection level is increased as the frequency detuning scales with $k$ and could overcome 25 GHz at higher order modes. The trade-off that provides proper decorrelation and efficient locking among all modes is a FSR difference of 5 GHz as shown in fig. 3. At this value the correlation coefficient of mode 1 and mode -1 is almost 0.5. BER is below $1 \times 10^{-3}$ for 5 GHz FSR detuning which makes our eight-mode RC system equivalent to a single-mode RC system with a delay of 1 ns, however with four-times lower speed penalty (6 instead of 25). In the same figure we can see as inset pictures the time traces of four significant modes (mode -1, 0, 1, 2) for FSR detuning equal to 0 and 9 GHz. For zero detuning the high correlation of the dynamical behavior of all modes is obvious.

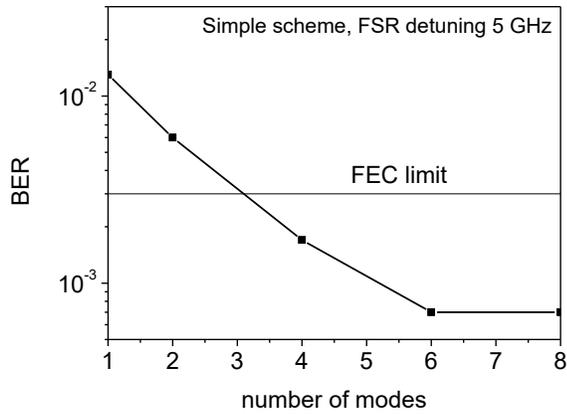

Fig. 4. BER performance as a function of the number of modes participating in the process (simple scheme, FSR detuning=5 GHz).

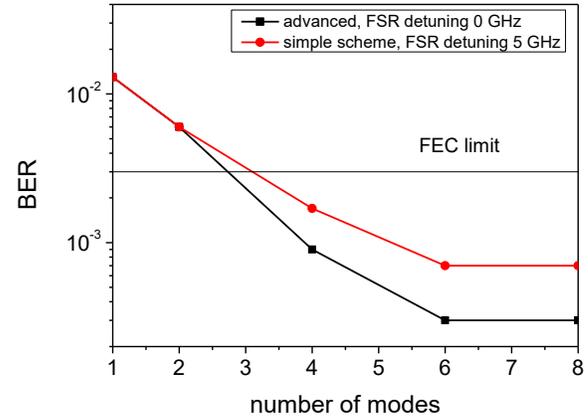

Fig. 5. BER performance as a function of the number of modes participating in the process (red line: simple scheme, FSR detuning=5 GHz, black line: advanced scheme, FSR detuning=0)

The processing power as a function of the number of modes participating in the output layer is depicted in fig. 4 for FSR detuning equal to 5 GHz, for which we obtained the best BER results. It is evident that as the number of modes increases (equivalently the number of virtual nodes increases as well), the BER performance improves till it reaches a plateau of $7 \times 10^{-4}$ for six modes. For eight modes the performance is identical, which means that the inclusion of higher order modes from this point on does not contribute to BER improvement. The underlying reasons of this behavior are the worse signal to noise ratio of higher order modes and partial correlation of the dynamics of the higher order modes with that of lower-order modes due to the fact that all modes share the same carrier pool which result in marginal improvement for more than 6 modes. In the next paragraph we study the behavior of the advanced system of fig. 1b.

### B. Evaluation of the advanced scheme

The advanced scheme of fig. 1b is a versatile platform that enables the modulation of each mode with a different mask, to increase the diversity among them and more importantly it offers the capability of performing multiple independent processing tasks at the same time. That is, different groups of modes may be employed for a different classification task. First we investigate the RC performance assuming that all modes are utilized in order to mitigate the dispersion of the same PAM-4 signal and we estimate the processing efficiency as a function of the slave modes that are taken into account in the linear regression. Since, each master mode enters slave laser with its own mask, the decorrelation at the input guarantees high decorrelation at the output, thus there is no need for introducing other decorrelation methods like FSR detuning presented in paragraph A. The simulations presented in fig. 5 consider that both lasers have the same FSR and that each master mode injects its slave counterpart with $\Delta\omega_k=0$. For moderate FSR detuning values (< 5 GHz), the BER performance was found to be unaffected by FSR detuning as expected. The results in fig. 5 clearly show that the BER plateau remains for the advanced scheme, however with a slightly improved performance ($3 \times 10^{-4}$) compared to the simple scheme ($7 \times 10^{-4}$), obviously due to better decorrelation offered when each mode is masked differently. What is important to study is the capability of the advanced system to process multiple signals at the same time. For instance, if we could send two separate tributaries of the same signal to two groups of modes, then we could double the processing power and thus reduce the speed penalty to 3. In order to better clarify this, let us assume that $L$ symbols of an impaired signal are detected. Then the signal could be split into two pieces of $L/2$ symbols and both pieces could be processed simultaneously by the same FP based RC system, simply by allocating each piece to a different group of processors (modes). In fig. 6 we can see the BER performance of the RC system when multiple PAM-4 signals are loaded to different group of modes (red line). For the sake of comparison, we have inserted the curve of fig. 5 which corresponds to applying one signal with different masks in all modes. In the case of processing two signals at the same time using four modes per each signal, we divide the eight modes in two groups, namely (-4, -2, 0, 2) and (-3, -1, 1, 3). Both groups provide almost identical BER performance (~ 0.0025). It is evident that parallel processing of multiple signals is characterized by degraded BER when compared to the case of loading the same signal in all modes. For instance, if we use only one signal at the input, the four modes would result in BER=$9 \times 10^{-4}$. This slight degradation is attributed to the inter-mode crosstalk due to the fact that all modes acquire gain from the same carrier population which affects the processing performance when different tasks are undertaken by different groups of nodes in analogy to cross-gain modulation effects in multi-channel amplification. Therefore, nonlinear crosstalk inside the slave laser limits parallel processing in only 2 simultaneous tasks as depicted in fig. 6. Despite this foreseen behaviour, degradation is not severe and the two signals are processed with final BER slightly below the FEC limit, which means that the FP RC has the ability to simultaneously "regenerate" two different PAM-4 signals in an efficient manner. Thus, speed penalty can be further improved by a factor of 2 and reach the record value of 3 or in other words this system can provide processing speed of 8 Gsa/s.

### C. Potential for photonic integration

Both systems depicted in fig. 1 have the potential to be miniaturized, even in the level of monolithic integration. The external cavity of slave laser can be monolithic integrated to the laser with advanced options for the control of feedback phase and amplitude as demonstrated in [22] and subsequent references. The lasers, modulators, variable attenuators, multiplexers/demultiplexers and photodiodes can be integrated with the use of III-V or hybrid III-V silicon photonics technologies [24-26]. The main difficulty is to integrate the circulator on chip although there has been recorded significant progress during the last decade in this field [27]. Even if one wants to avoid to exploit a technology which is still immature for the on-chip development of optical isolators/circulators, the system could still remain miniaturized by combining two chips (chip1: master laser and modulation stages, chip2: slave laser, external cavity, photodetection and electronics) with a miniaturized fiber-based isolator. The excess losses are estimated to be 6 dB if grating couplers are used at both ends and polarization maintaining fiber must be used to avoid polarization rotations. Compensation for the excess coupling losses and further control of injection strength could be obtained with the use of a semiconductor amplifier placed either after master laser or before slave laser. Such a device is practical and not sensitive in external mechanical or temperature vibrations.

### IV. CONCLUSIONS

This paper presented the potential of off-the-shelf Fabry Perot lasers to constitute powerful reservoir computing machines offering parallelism in mode granularity. Extensive simulations showed that up to eight modes can be leveraged for parallel processing and demonstrated that almost real-time mitigation of fiber transmission impairments in 25 Gbaud signals is possible, reducing the speed penalty to 3 and prove the potential for 8 Gsa/s processing speed. The work also showed that different

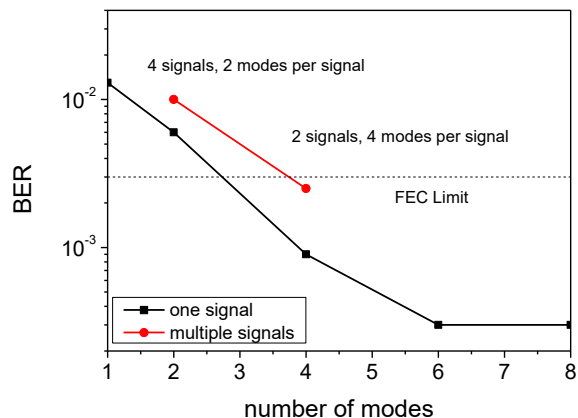

Fig. 6. BER performance of the advanced system as a function of the number of modes participating in the process considering one signal allocated to all modes (black line) and more than one signals processed concurrently by allocating them to different groups of modes (red line)

tasks can be run in different sets of modes with acceptable cross-talk, thus FP based processors can compensate for the dispersion of more than one signals at the same time. Finally, the potential for miniaturization was discussed showing that the proposed system can be considered as a practical and powerful solution for photonic neuromorphic computing.

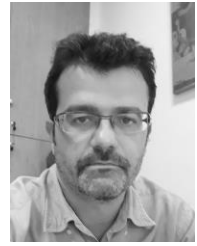

**Adonis Bogris** was born in Athens. He received the B.S. degree in informatics, the M.Sc. degree in telecommunications, and the Ph.D. degree from the National and Kapodistrian University of Athens, Athens, in 1997, 1999, and 2005, respectively. His doctoral thesis was on all-optical processing by means of fiber-based devices. He is currently a Professor at the Department of Informatics and Computer Engineering at the University of West Attica, Greece. He has authored or co-authored more than 150 articles published in international scientific journals and conference proceedings and he has participated in plethora of EU and national research projects. His current research interests include high-speed all-optical transmission systems and networks, nonlinear effects in optical fibers, all-optical signal processing, neuromorphic photonics, mid-infrared photonics and cryptography at the physical layer. Dr. Bogris serves as a reviewer for the journals of the IEEE and OSA.

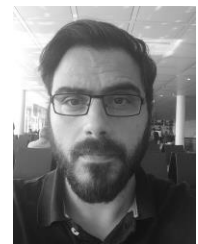

**Charis Mesaritakis** received his BS degree in Informatics, from the department of Informatics & Telecommunications of the National & Kapodistrian University of Athens in 2004. He received the MSc in Microelectronics from the same department, whereas in 2011 he received his Ph.D degree on the field of quantum dot devices and systems for next generation optical networks, in the photonics technology & optical communication laboratory of the same institution. In 2012 he was awarded a European scholarship for post-doctoral studies (Marie Curie FP7-PEOPLE IEF) in the joint research facilities of Alcatel-Thales-Lucent in Paris-France, where he worked on intra-satellite communications. He has actively participated as research engineer/technical supervisor in more than 10 EU-funded research programs (FP6-FP7-H2020) targeting excellence in the field of photonic neuromorphic computing, cyber-physical security and photonic integration. He is currently an Associate Professor at the Department of Information & Communication Systems Engineering at the University of the Aegean, Greece. He is the author and co-author of more than 60 papers in highly cited peer reviewed international journals and conferences, two international book chapters, whereas he serves as a regular reviewer for IEEE and OSA.

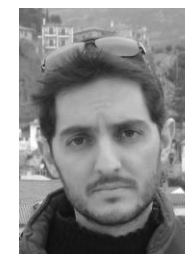

**Stavros Deligiannidis** (holds a BSc in Physics, a MSc degree in Microelectronics and VLSI from the National and Kapodistrian University of Athens. Since 2010 he is with the Department of Computer Engineering of the Technological Educational Institute of Peloponesse, Greece where he serves as a Lecturer. He is currently pursuing his PhD degree at the University of West Attica in the field of novel signal processing techniques for optical communication systems under the supervision of Prof. Adonis Bogris. He has worked as a researcher in local and European projects. His current research interests include optical communications, deep learning, digital signal processing, and parallel computing.

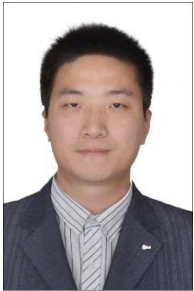

**Pu Li** received the M.S. degree in physical electronics from Taiyuan University of Technology (TYUT), Shanxi, China, in 2011, and the Ph.D. degree in circuits and systems from the Key Laboratory of Advanced Transducers and Intelligent Control System (Ministry of Education of China), TYUT, in 2014.

In 2014, he joined TYUT as a lecturer. He was a Visiting Scholar in the School of Electronic Engineering, Bangor University, U.K., in 2017. Since 2018, he is a Professor in the Key Laboratory of Advanced Transducers and Intelligent Control System (Ministry of Education of China), TYUT. His research interests include nonlinear dynamics of semiconductor lasers and its applications, and all-optical analog-to-digital conversion.

Dr. Li serves as a Reviewer for journals of the IEEE, OSA, and Elsevier organizations.